\documentclass[%
 aip,
 jmp,%
 amsmath,amssymb,
 reprint,%
]{revtex4-1}
\usepackage{graphicx}
\usepackage{psfig}
\usepackage{dcolumn}
\usepackage{bm}
\usepackage{color}

\newcommand{\beq}{\begin{equation}}
\newcommand{\eeq}{\end{equation}}
\newcommand{\ba}[1]{\begin{array}{#1}}
\newcommand{\ea}{\end{array}}
\newcommand{\bea}{\begin{eqnarray}}
\newcommand{\eea}{\end{eqnarray}}
\newcommand{\nn}{\nonumber \\}
\newcommand{\ben}{\begin{enumerate}}
\newcommand{\een}{\end{enumerate}}
\newcommand{\bit}{\begin{itemize}}
\newcommand{\eit}{\end{itemize}}
\newcommand{\bde}{\begin{description}}
\newcommand{\ede}{\end{description}}

\newcommand{\ds}{\displaystyle}

\newcommand{\sz}{\scriptsize}

\newcommand{\CANVIS}[1]{\textcolor{black}{#1}}

\begin{document}

\title{Mesoscopic analysis of networks: applications to exploratory analysis and data clustering} 

\author{Clara Granell}
\noaffiliation
\author{Sergio G\'omez}
\noaffiliation
\author{Alex Arenas}
\email{alex.arenas@urv.cat}
\noaffiliation
\affiliation{Departament d'Enginyeria Inform\`atica i Matem\`atiques, Universitat Rovira i Virgili, 43007 Tarragona, Catalonia, Spain}

\date{\today}

\begin{abstract}
We investigate the adaptation and performance of modularity-based algorithms, designed in the scope of complex networks, to analyze the mesoscopic structure of correlation matrices. Using a multi-resolution analysis we are able to describe the structure of the data in terms of clusters at different topological levels. We demonstrate the applicability of our findings in two different scenarios: to analyze the neural connectivity of the nematode {\em Caenorhabditis elegans}, and to automatically classify a typical benchmark of unsupervised clustering, the Iris data set, with considerable success.
\end{abstract}


\maketitle 

\begin{quotation}

Facing the famous Salvador Dali's painting ``Gala contemplating the Mediterranean sea which at twenty meters becomes a portrait of Abraham Lincoln", we have the best proof of how a complex system reveals different information when observed at different (in this case length) scales. We proposed a method \cite{njp08} to unveil the equivalent phenomena in the description of complex networks from a topological perspective. By defining a parameter that controls the resistance of each node to belong to a group, we are able to analyze the community structure of the network at different topological scales. We apply the method to the exploratory analysis of the structural connectivity of the neuronal system of {\em C.~elegans} and find a tentative classification of functional activity of groups of neurons at certain topological scales. We also have tested the method to automatically classify a typical benchmark of unsupervised data clustering, the Iris dataset. These results pave the way to the applicability of community detection algorithms in complex networks to the exploration and classification of real data sets.

\end{quotation}

\section{Introduction}

Complex networks are graphs representative of the intricate connections between elements in many natural and artificial systems\cite{strogatz,havlin,barabasi}, whose description in terms of statistical properties has been largely developed in the curse for a universal classification of them. However, when the networks are locally analyzed some characteristics that become partially hidden in the statistical description emerge. The most relevant perhaps is the discovery in many of them of {\em community structure}, meaning the existence of densely (or strongly) connected groups of nodes, with sparse (or weak) connections between them\cite{firstnewman}.

The study of the community structure  helps to elucidate the organization of the networks and, eventually, could be related to the functionality of groups of nodes\cite{amaral}. The most successful solutions to the community detection problem, in terms of accuracy, are those based in the optimization of a quality function called {\em modularity} proposed by Newman and Girvan\cite{newgirvan} that allows the comparison of different partitioning of the network. Given a network partitioned into communities, being $C_i$ the community to which node $i$ is assigned, the mathematical definition of modularity is expressed in terms of the weighted adjacency matrix $w_{ij}$, that represents the value of the weight in the link between nodes $i$ and $j$, this weight would be $0$ if no link existed, and the strengths $ w_i=\sum_j w_{ij}$ as \cite{newanaly}
\begin{equation}
Q=\frac{1}{2w}\sum_i \sum_j \left( w_{ij}-\frac{w_i w_j}{2w}\right)\delta(C_i,C_j)\,,
\label{QW}
\end{equation}
\noindent where the Kronecker delta function $\delta(C_i,C_j)$ takes the values, 1 if node $i$ and $j$ are into the same community,  0 otherwise, and the total strength $2w=\sum_i w_i$.
The modularity of a given partition is then, the probability of having  edges falling within groups in the network minus the expected probability in an equivalent (null case) network with the same number of nodes, and edges placed at random preserving the nodes' strength. The larger the modularity the best the partitioning is, cause more deviates from the null case. Note that the optimization of the modularity cannot be performed by exhaustive search since the number of different partitions is equal to the Bell\cite{bell} or exponential numbers, which grow at least exponentially in the number of nodes $N$. Indeed, optimization of modularity is a NP-hard (Non-deterministic Polynomial-time hard) problem\cite{brandes}. Several authors have attacked the problem, with considerable success, by proposing different optimization heuristics\cite{newfast, clauset, rogernat, duch, pujol, newspect}, see Fortunato\cite{fortunato_rev} for a review.

Maximizing modularity one obtains the ``best" partition of the network into communities. This partition represents an intermediate topological scale of organization, or mesoscale, that in many cases has been shown to coincide with known information about subdivisions in the network\cite{newgirvan,jstat}. However, recently, it has been pointed out that the optimization of the modularity has a characteristic scale related to the number of links in the network, that delimits the resolution beyond which no separation into smaller groups can be obtained when optimizing modularity, even-though these smaller partitions, and then different levels of description, are plausible to exist from direct observation\cite{fortunato}. The problem seems then that modularity, as it has been prescribed, does not have access to these other levels of description, \CANVIS{and then its direct interpretation must be cautiously used\cite{good}}. The reason for this is that the topological scale at which we have access by maximizing modularity has a topological resolution limit. The analogy with the observation of Dali's painting is clear, modularity is our tool to ``observe" a complex network, and their limit is equivalent of a limit in the distance at which we observe the painting (Fig.~1).
\begin{figure}
\centering
\resizebox{1.0\columnwidth}{!}{
\includegraphics{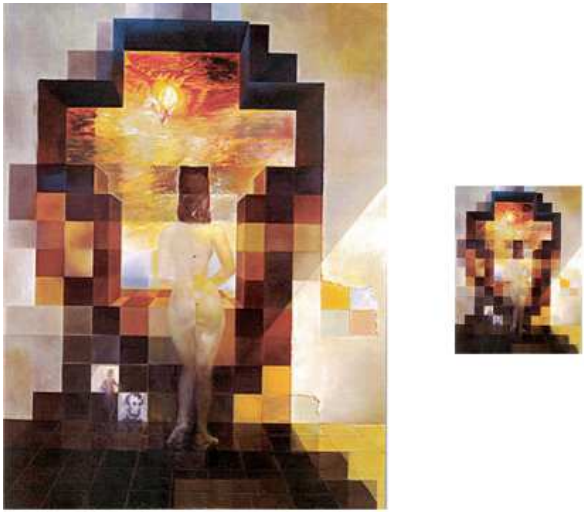}
}
\caption{{\em ``Gala contemplating the Mediterranean sea which at twenty meters becomes a portrait of Abraham Lincoln"}, by Salvador Dali, 1974.
Left, at closer distance, and right, at larger distance.}
\label{fig:dali}
\end{figure}
We proposed a method\cite{njp08} that allows the full screening of the topological structure at any resolution level using the original formulation and semantics of modularity, overcoming then the resolution limit. Our aim is to take advantage of this method to analyze real data sets in terms of clustering.

The paper is structured as follows: In the next section we overview the multiple resolution method. Once the method has been presented, we propose its application for exploratory analysis in the topology of the neural network of the nematode {\em C.~elegans} in section III, and its application to data clustering in section IV. Finally we present the conclusions of the work in section V.

\section{Multiple resolution method}

\CANVIS{In this section we provide the necessary tools to extend the multiple resolution method to the most general case of networks with weighted signed directed links.}


\subsection{General formulation of modularity}

The generalization of modularity to any network, with weighted, directed and signed values of the weights\cite{signed} is as follows.
Let us suppose that we have a weighted undirected complex network with weights $w_{ij}$
as above. The relative strength $p_i$ of a node
\beq
  p_i = \frac{w_i}{2 w}\,,
\eeq
may be interpreted as the probability that this node makes links to other
ones, if the network were random. This is precisely the approach taken
by Newman and Girvan to define the modularity null case term, which reads
\beq
  p_i p_j = \frac{w_i w_j}{(2 w)^2}\,.
\eeq

The introduction of negative weights destroys this probabilistic interpretation
of $p_i$, since in this case the values of $p_i$ are not guaranteed to be between
zero and one. The problem is the implicit hypothesis that there is only one
unique probability to link nodes, which involves both positive and negative
weights. To solve this problem, we have to introduce two different probabilities to
form links, one for positive and the other for negative links.

Let us formalize this approach. First, we separate the positive and negative
weights:
\beq
  w_{ij} = w_{ij}^{+} - w_{ij}^{-}\,,
\eeq
where \CANVIS{we use the notation}
\bea
  w_{ij}^{+} & = & \max\{0,  w_{ij}\}\,, \\
  w_{ij}^{-} & = & \max\{0, -w_{ij}\}\,.
\eea
\CANVIS{These expressions are useful since in principle we do not know the sign of $w_{ij}$.}
The positive and negative strengths are given by
\bea
  w_i^{+} & = & \sum_j w_{ij}^{+}\,, \\
  w_i^{-} & = & \sum_j w_{ij}^{-}\,,
\eea
and the positive and negative total strengths by
\bea
  2 w^{+} & = & \sum_i w_i^{+} = \sum_i \sum_j w_{ij}^{+}\,, \\
  2 w^{-} & = & \sum_i w_i^{-} = \sum_i \sum_j w_{ij}^{-}\,.
\eea
Consequently,
\beq
  w_i = w_i^{+} - w_i^{-}
\eeq
and
\beq
  2 w = 2 w^{+} - 2 w^{-}\,.
\eeq

With these definitions at hand, the connection probabilities with
positive and negative weights are respectively
\bea
  p_i^{+} & = & \frac{w_i^{+}}{2 w^{+}}\,, \\
  p_i^{-} & = & \frac{w_i^{-}}{2 w^{-}}\,.
\eea

Now, there are two terms which contribute to modularity: the first
one takes into account the deviation of actual positive weights
against a null case random network given by probabilities $p_i^{+}$,
and the other is its counterpart for negative weights. Thus, it
is useful to define
\bea
  Q^{+} & = & \frac{1}{2w^{+}} \sum_i \sum_j \left(
                w_{ij}^{+} - \frac{w_i^{+} w_j^{+}}{2w^{+}}
              \right) \delta(C_i,C_j)\,,
  \\
  Q^{-} & = & \frac{1}{2w^{-}} \sum_i \sum_j \left(
                w_{ij}^{-} - \frac{w_i^{-} w_j^{-}}{2w^{-}}
              \right) \delta(C_i,C_j)\,.
\eea

The total modularity must be a trade off between the tendency of
positive weights to form communities and that of negative weights
to destroy them. If we want that $Q^{+}$ and $Q^{-}$ contribute
to modularity proportionally to their respective positive and negative
strengths, the final expression for modularity $Q$ is
\beq
  Q = \frac{2 w^{+}}{2 w^{+} + 2 w^{-}} Q^{+} -
      \frac{2 w^{-}}{2 w^{+} + 2 w^{-}} Q^{-}\,.
\eeq
An alternative equivalent form for modularity $Q$ is
\bea
  Q = \frac{1}{2w^{+} + 2 w^{-}} & \ds \sum_i \sum_j &
        \left[ w_{ij} - \left(
          \frac{w_i^{+} w_j^{+}}{2w^{+}}- \frac{w_i^{-} w_j^{-}}{2w^{-}}
        \right) \right] \nn
        &&\times \delta(C_i,C_j)\,.
  \label{QWS}
\eea

The main properties of Eq.~(\ref{QWS}) are the following: without negative weights, the
standard modularity is recovered; modularity is zero when all nodes are together
in one community; and it is antisymmetric in the weights, i.e.
$Q(C,\{w_{ij}\}) = - Q(C,\{-w_{ij}\})$\,.

The extension to directed networks\cite{njp07} is simply obtained by the substitutions
in Eq.~(\ref{QWS}) of
\bea
  w_{i}^{\pm} &\rightarrow &  w_{i}^{\pm,\mbox{\scriptsize out}} = \sum_{k} w^{\pm}_{ik}\,,\\
  w_{j}^{\pm} &\rightarrow &  w_{j}^{\pm,\mbox{\scriptsize in}} = \sum_{k} w^{\pm}_{kj} \,.
\eea

\subsection{\CANVIS{Mesocales analysis for weighted signed networks}}

\CANVIS{
The extension of the multiple resolution method\cite{njp08} to the general case of weighted signed networks follows the same original idea. The method relies on the introduction of a magnitude $r$ that we call {\em resistance}, represented by a self-link for each node, that stands for the opposition of a node to belong to a group, in the sense of modularity. We tune the resistance uniformly for all nodes because in this way the functional form of the strength distribution is preserved and does not distort the relative structural properties of nodes. More precisely, the formulation of modularity $Q_r$ at different resolution scales tagged by $r$ consists in substituting in Eq.~(\ref{QWS})
\bea
  w_{ij} & \rightarrow & w_{ij} + r \delta_{ij} \,,\\
  w_{i}^{\pm} & \rightarrow & w_{i}^{\pm} + r^{\pm} \,, \\
  2w^{\pm} & \rightarrow & 2w^{\pm} + Nr^{\pm} \,,
\eea
\noindent where
\beq
  r = r^{+} - r^{-}\,,
\eeq
and
\bea
  r^{+} & = & \max\{0,  r\}\,, \\
  r^{-} & = & \max\{0, -r\}\,.
\eea
}

\CANVIS{
The topological scale determined by maximizing $Q$ at which the detection of community structure has been attacked so far, corresponds to $r=0$ (Newman's scale). For positive values of $r$, we have access to the substructure below $r=0$, and for negative values of $r$ we have access to the superstructures. For negative values of $r$, the resistance should be understood as an affinity of nodes to belong to the same group, and using Eq.~(\ref{QW}) the formulation is still preserved but not the semantics in terms of probabilities. The main challenge in this new scenario is that the limiting cases of $r$ that corresponds to the partition of individual nodes, and to the whole network as a unique module have to be computed using the new modularity formulation Eq.~(\ref{QWS}).
}

\subsection{Resistance limiting cases for weighted signed networks}

\CANVIS{
Here we present the mathematical proofs of the physical limiting cases of the resistance for weighted signed networks. Let us call $r_{\max}$ the limit of resistance for which all nodes are isolated in communities of size~1, and $r_{\min}$ the limit for which all nodes become members of a single group that represents the whole network.} To determine $r_{\max}$ we look for a value of the resistance such that the increment in modularity when joining any pair of vertices in the same community is negative, and the contrary for $r_{\min}$. \CANVIS{The idea is the following: if $r>0$ and all the non-diagonal terms ($i\ne j$) of Eq.~(\ref{QWS}) are negative,
\beq
  w_{ij} \le \frac{(w_i^{+}+r) (w_j^{+}+r)}{2w^{+}+N r} - \frac{w_i^{-} w_j^{-}}{2w^{-}} \,,
  \ \ \forall i \ne j\,,
  \label{rmaxeqs}
\eeq
then the maximum of $Q_r$ is achieved with the partition which satisfies $\delta(C_i,C_j)=0$ for all $i\ne j$, i.e.\ the partition in which all nodes are isolated. Eqs.~(\ref{rmaxeqs}) form a system of second order inequations in $r$. After some algebra, it can be shown that $r_{\max}$ is the lowest value of $r$ for which the following set of inequalities per link (denoted $ij$) is satisfied:}
\beq
\min_{r,ij} [Ar^2+B_{ij}r+C_{ij} \le 0]
\label{rmax}
\eeq
where
\bea
A &=& -2w^-\\
B_{ij} &=& N(2w^-w_{ij} + w_{i}^{-} w_{j}^{-}) - 2w^-(w_{i}^{+}+w_{j}^{+})\\
C_{ij} &=&  2w^{-}2w^{+}w_{ij} + 2w^{+}w_{i}^{-}w_{j}^{-} - 2w^{-}w_{i}^{+}w_{j}^{+}\
\eea

\CANVIS{
Equivalently, if $r<0$ and all the non-diagonal terms ($i\ne j$) of Eq.~(\ref{QWS}) are positive,
\beq
  w_{ij} \ge \frac{w_i^{+} w_j^{+}}{2w^{+}} - \frac{(w_i^{-}-r) (w_j^{-}-r)}{2w^{-}-N r} \,,
  \ \ \forall i \ne j\,,
  \label{rmineqs}
\eeq
the maximum of $Q_r$ is achieved with the partition which satisfies $\delta(C_i,C_j)=1$ for all $i\ne j$, i.e.\ the partition in which all nodes are together in the same community.} Thus, to determine a lower bound of $r_{\min}$ we look for the largest value of $r$ satisfying
\beq
\max_{r,ij} [Ar^2+B_{ij}r+C_{ij} \ge 0]
\label{rmin}
\eeq
where
\bea
A &=& 2w^+\\
B_{ij} &=& N(2w^+w_{ij} -w_{i}^{+} w_{j}^{+}) +2w^+(w_{i}^{-}+w_{j}^{-})\\
C_{ij} &=&  2w^{+}2w^{-}w_{ij} - 2w^{-}w_{i}^{+}w_{j}^{+} + 2w^{+}w_{i}^{-}w_{j}^{-}\
\eea

The value of $r$ obtained from Eqs.~(\ref{rmin}) is only a lower bound of the exact $r_{\min}$, since these equations are only sufficient conditions for the existence of a unique communty holding all the nodes of the network (not all terms in Eq.~(\ref{QWS}) need to be positive in the $r_{\min}$ limit). On the other hand, Eqs.~(\ref{rmax}) are necessary and sufficient conditions, and thus the $r_{\max}$ found is the exact value.

The method to unveil the mesoscales of a complex network consists in to optimize $Q_r$ for $r$ in $[r_{\min}, r_{\max}]$. Different values of $r$ will eventually reveal different optimal partitions (found by heuristic algorithms to detect community structure) that represent intermediate topological scales of the complex network.
We have applied this method to study the mesoscales in synthetic structured networks and real complex networks.

\subsection{Validation of the method in synthetic networks}

\begin{figure}
  \begin{center}
   \mbox{\psfig{file=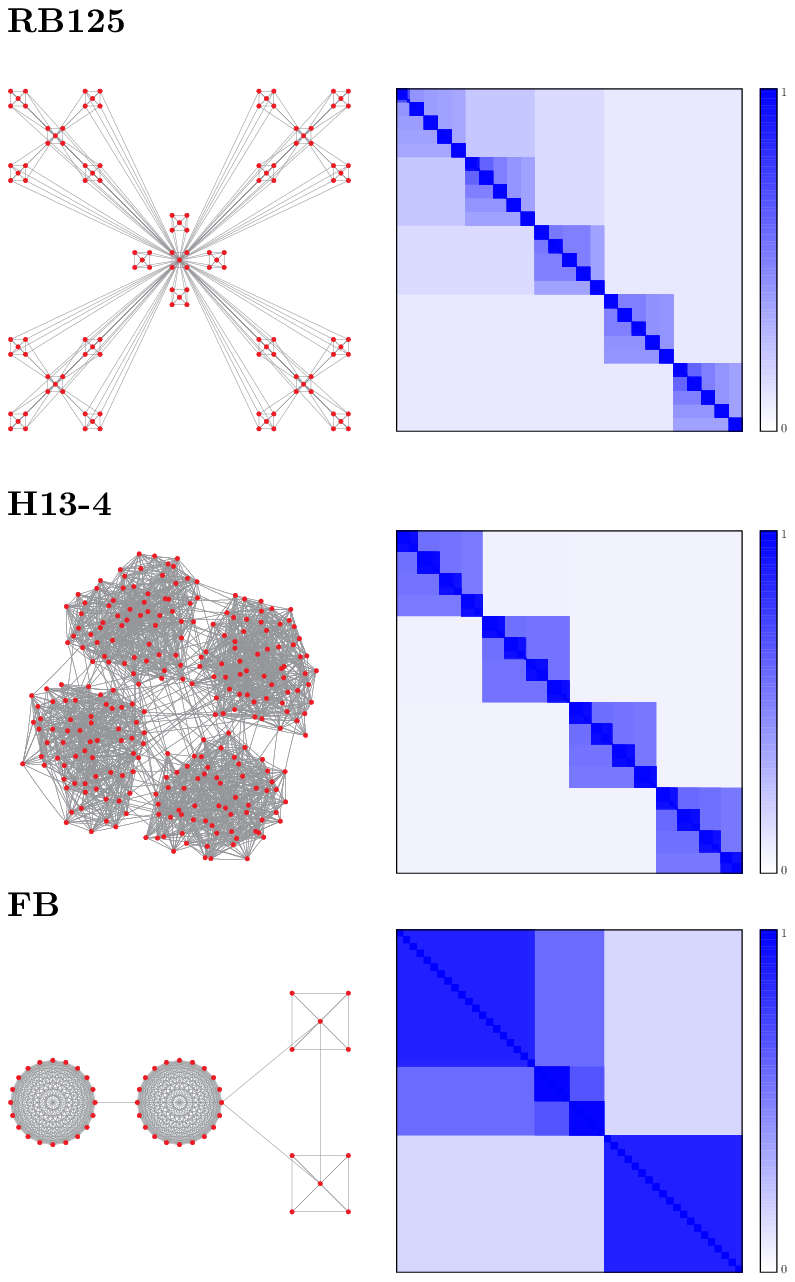,width=1.0\columnwidth}}
  \end{center}
\caption{\CANVIS{Frequency} mesoscales matrices in synthetic complex networks. We have computed the topological mesoscales for three synthetic networks. Left, we plot the networks and right we present their mesoscales matrices.
The different color levels correspond to the superposition of the structures in $r$, which account for the persistence of the partitions revealed. See text for details.
}
\label{fig:meso}
\end{figure}

In Fig.~\ref{fig:meso} we have screened the whole range of topological scales for three synthetic networks, representing the number of modules obtained at the optimal partition for $Q_r$ and plotting in a matrix the superposition of scales found.  \CANVIS{More precisely, any graphical representation of the whole mesoscale should take into account, for every pair of nodes, the frequency of mesoscales at which they belong to the same community. Each mesoscale has a natural {\em length} defined by the range of resistances $[r_{\mathrm{from}},r_{\mathrm{to}}]$ at which it is optimal:
\begin{equation}
  \mbox{length} = \log(r_{\mathrm{to}}-r_{\mathrm{min}}) - \log(r_{\mathrm{from}}-r_{\mathrm{min}}) \enspace .
\end{equation}
Thus, the length frequency for a pair of nodes is the sum of the lengths corresponding to mesoscales in which they belong to the same community, normalized by the total length. The graphical representation of this table is the {\em frequency mesoscales matrix}.}
First we have computed the modular structure in a hierarchical scale-free network with 125 nodes, RB~125, proposed by Ravasz and Barabasi\cite{RB}. We clearly observe persistent structures in 5 and 25 communities respectively, that account for the subdivisions more significant in the process, showing two hierarchical levels for the structure.

Another network example used is the H~13-4 network\cite{arenas}, which corresponds to a homogeneous in degree network with two predefined hierarchical levels, being 256 the number of nodes, 13 the number of links of each node with the most internal community (formed by 16 nodes), 4 the number of links with the most external community (four groups of 64 nodes), and 1 more link with any other node at random in the network. Both hierarchical levels are revealed by the method as they correspond to the original construction of the network: the first hierarchical level consisting in 4 groups of 64 nodes, and the second level consisting in 16 groups of 16 nodes.

Finally, we have used the FB network proposed by Fortunato and Barth{\'e}lemy\cite{fortunato} to demonstrate the resolution limit of modularity (at $r=0$). It consists in two cliques of 20 nodes linked with two small cliques of 5 nodes. At $r=0$ the best partition cannot separate the two small cliques. We observe that the partition searched by the authors, formed by the four cliques isolated in their own communities, is obtained by increasing the resolution $r$, showing that the resolution limit of modularity is overcome by the method.

The optimization of modularity in all these cases has been performed using existing heuristics found in the literature\cite{duch,njp08,newspect} and compiled in a free toolbox available at the authors' webpage\cite{gomez}.

\section{Application to exploratory data analysis}

\begin{figure}[t]
  \begin{center}
  \mbox{\psfig{file=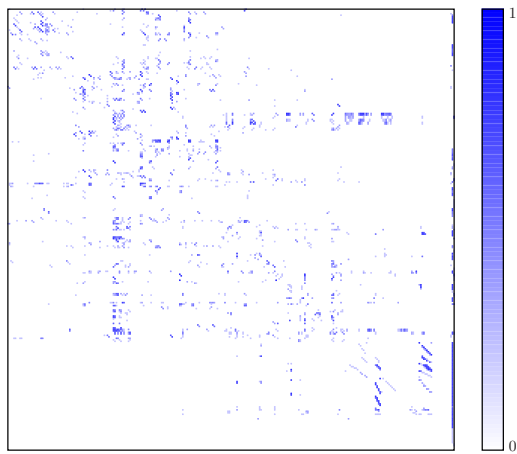,width=6.8cm}}
  \end{center}
  \caption{Connectivity matrix of {\em C.~elegans} neuronal network.}
\label{CEmatrix}
\end{figure}

Exploratory data analysis stands for the approach to data analysis in which some rather general assumptions are used to reveal information
of the data in a kind of inverse hypothesis testing. In our particular scenario, we will analyze the structure of the neural connectivity of the nematode {\em C.~elegans}\cite{white} using this approach. \CANVIS{We do not pretend an exhaustive biological classification of all functionalities that are related to the topology but to show the applicability of the mesoscales analysis described before.
A pretty exhaustive analysis of the same system has been recently presented\cite{raj} for the scale corresponding to $r=0$.}
The whole nervous system of the nematode is composed by 302 neurons whose anatomical and connectivity description is completely known. The resulting network is represented as a weighted directed adjacency matrix, see Fig.~\ref{CEmatrix}. We will assume that those groups of nodes more persistent throughout the screening of the mesoscales of the topology have some functional role, and after we will look for this role in the current biological literature.

The original data\cite{stro} is a weighted and directed network, composed of 306 vertices (302 neurons + WE, WI, WM and WN) and 2359 arcs. We have discarded nine disconnected nodes from the network, the remaining 297 neurons form a single connected component and will be the subject of our analysis.

We have discretized the resistance range in 1000 non-uniform intervals, in such a way that the last resistance increment is ten times larger than the first one, and the size of the increments grow at a constant rate. The significant Newman's scale $r=0$ has been added. The negative values of the resistance have been discarded, since we are interested only in sub-structure beyond the standard Newman's scale\cite{lncs}.

\begin{figure}[t]
  \begin{center}
  \mbox{\psfig{file=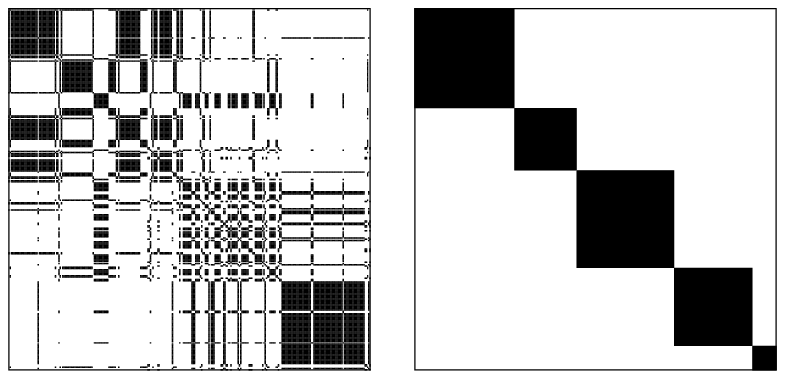,width=1.0\columnwidth}}
  \end{center}
  \caption{Newman's scale of the {\em C.~elegans} neuronal network. Left, original order, right, reordering by communities.}
  \label{rcero}
\end{figure}
The order of the neurons in the matrix follows that in Watts and Strogatz\cite{stro} obtained from experimental data by White~et~al.\cite{white}. The detection of the mesoscales in this neuronal system has been performed according to the method explained in the previous section. The best partition at $r=0$ corresponding to the original Newman's scale provides with 5 communities. The representation of the obtained groups is depicted in Fig.~\ref{rcero} (left).
This figure does not allow the observation of relevant information because the original order of the neurons in Fig.~\ref{CEmatrix}, however after ordering the neurons in the matrix by their communities, the representation shown in Fig.~\ref{rcero} emerges.

\begin{figure}[tcb]
  \begin{center}
  \mbox{\psfig{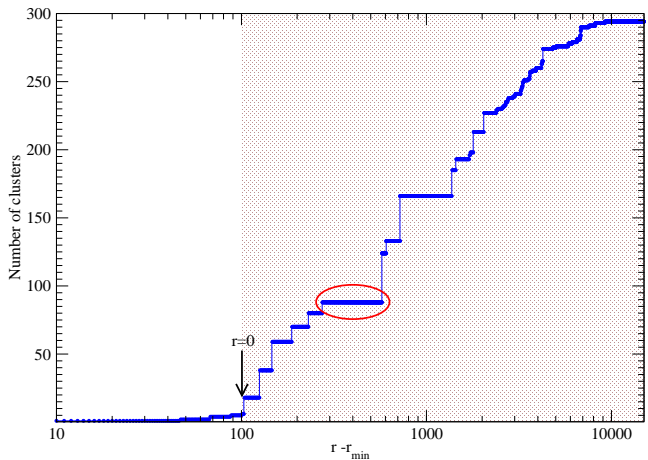}}
  \end{center}
  \caption{\CANVIS{Mesoscales of the C.~elegans: number of clusters in the optimal partition at every value of the topological scale defined by the $\log(r-r_{\mbox{\sz min}})$, where $r_{\mbox{\sz min}}$ refers to the exact value, not its lower bound. Highlighted in circle, we represent the scale that most contributes to the frequency matrix.}}
\label{meso_cele}
\end{figure}

The coarse graining at $r=0$ provides then with a large scale in this system, hence our interest has been specially focused in the sub-structural levels, \CANVIS{not in supra-structural levels}, that means that we have analyzed the mesoscale for $r \in [0,r_{\mbox{\sz max}}]$, \CANVIS{see gray region of  Fig.~\ref{meso_cele}}. We used the partition at $r=0$ simply as a reference for sorting the neurons in the substructures found by the multiple resolution method.

Any trial of classification of the functional role of neurons of the {\em C.~elegans} is extremely delicate because the multifunctional aspects they have. Many neurons participate in different synaptic pathways resulting in different functionalities. This property is also captured by our method that shows that at different scales the same neuron can appear in different groups, i.e. the method is not necessarily hierarchical. However, to extract information from the results obtained, we use an ensemble of the different partitions found by screening $r$, and construct a \CANVIS{frequency} mesoscales matrix, indicating the relative persistence of each neuron in a particular community. By fixing a threshold in the frequency value, we are able to unravel sub-structural scales that correspond to groups of neurons involved in different functionalities at different time scales.
\begin{figure}[tcb]
  \begin{center}
  \mbox{\psfig{file=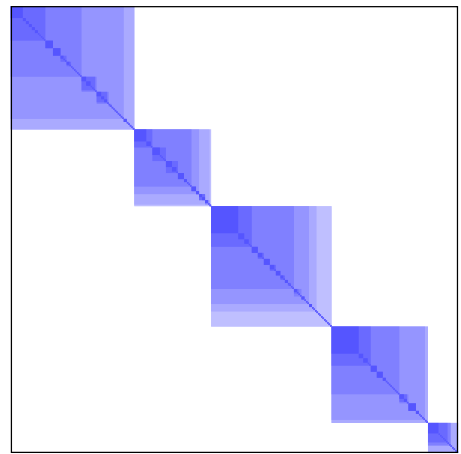,width=6cm}}
  \end{center}
  \caption{Frequency matrix of {\em C.~elegans} neuronal network thresholded at 0.6. We used a color scale (same as in Fig.\ref{CEmatrix}) to plot the persistence of neurons into the same groups, darker values corresponds to more persistent communities and, according to our hypothesis in the exploratory analysis, to specific functionalities}
\label{freqtr}
\end{figure}

The most interesting information is that provided at a large value of the frequency threshold, because in this case the substructures found will contain small groups of neurons whose activity response is topologically correlated, \CANVIS{in particular the highlighted scales in Fig.~\ref{meso_cele} are the ones that most contribute to the frequency matrix}. We have studied the ensemble frequency matrix at a threshold value of $0.6$, Fig.~\ref{freqtr}, \CANVIS{the lengths below the threshold are discarded, and the connected components of the graph defined by the remaining lengths are found}. We have chosen this threshold fixing the sizes of the groups to be analyzed to be less than ten neurons. With this information at hand, and the wide description of each neuron found at the public database of {\em C.~elegans}\cite{durbin,wa}, we propose a tentative classification of some groups of neurons by functionality.

Our purpose, after identification of individual functionalities, has been to assign a specific action to the more persistent groups of neurons. The classification obtained (see appendix) does not pretend to be exact but to provide biologists with a useful information for future research.

\section{Application to the unsupervised classification of data}

Unsupervised classification of data (or data clustering) stands for the process of grouping patterns of data according to their similarity. A pattern is a vector of features (usually understood as a point in a multidimensional space) that describes the item we wish to classify. The goal of the process of data clustering is to organize these patterns into groups, in such a way that patterns into the same group are more alike than with other patterns in other groups.

The problem of data clustering has been the subject of interest in many disciplines where the mining of raw information is crucial to understand some phenomenon or gain insight into a system. Typical processes where data clustering is used are pattern analysis, decision-making, machine learning and image segmentation. These subjects have interesting applications as for example targeted marketing, biological taxonomy and detecting communities of interest in the World Wide Web\cite{clust_app}.

The methodology used to obtain the clusters from the raw data is as follows: First of all, a representation of the patterns has to be chosen, and also a feature selection or extraction is performed. Feature selection means choosing, from all the available features, those that will make easier the process of clustering, leaving the redundant, correlated and less informative features out of the analysis. On the other hand, feature extraction consists in transforming the original dataset to a new one containing only the most relevant information. This first step is very important, as the result of the clustering often depends directly of the quality of it. Secondly, the similarity or dissimilarity between each pair of patterns has to be computed, which is often done by defining a measure of distance. The result of this step is the similarity matrix, which using the mapping to complex networks can be understood as a graph, where each node is a pattern and the links are the representation of the similarity\cite{rev_dc}. Finally, the main step of the process, the grouping (or clustering) algorithm, which will decompose the similarity matrix and return the groups of data.

\begin{figure}
\centering
\resizebox{1.0\columnwidth}{!}{%
  \includegraphics*{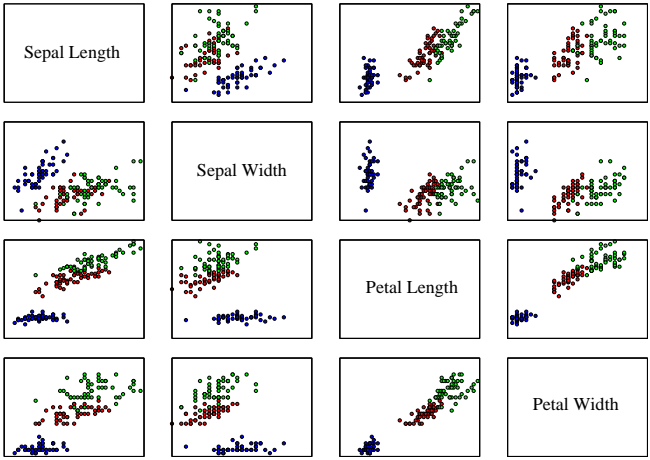}
}
\caption{Feature vectors for the Iris data set. Colors correspondence are: setosa-blue, versicolor-red, and virginica-green.}
\label{iriscol}
\end{figure}

In our approach, the algorithm used to classify the similarity matrix is the multiple resolution algorithm based on modularity explained previously in this document. Given the nature of this algorithm, the result will not be a single partition into clusters, but a collection of different partitions. This fact deserves a reflection about how to evaluate the quality of the output obtained. If we make a screening between the minimum and maximum value of the resistance parameter to obtain every topological scale of resolution of the network, each one of these resolution levels will provide us with a partition of clusters. Then the question is, which one of these partitions is the right one? The answer is that every one of them is right, since what we are doing is analyzing the network at different levels of resolution, and all the information obtained though this process is found in the structure of the network. Having pointed that out, the problem of choosing the right partition is translated to that of choosing the more relevant partitions. The more relevant partitions in our scope are those that persist unchanged during larger intervals of values of the resistance parameter.

The dataset benchmark selected to perform the data clustering is the Iris flower dataset, presented by Sir Ronald Aylmer Fisher\cite{fisher} in 1936. This dataset consists of 150 patterns corresponding to three different classes of flowers: Setosa, Versicolor and Virginica. Four features, the width and length of petal and sepal, form each pattern. Plots for the cross-variables and type of flowers are represented in Fig.~\ref{iriscol}. The unsupervised classification of this dataset is a major challenge in artificial intelligence and statistical theory, because of the patterns' organization, while one of the classes is linearly separable and then easily to classify by any elemental classification algorithm, the other two classes are not linearly separable and consequently far more difficult to classify.

\begin{figure}
\centering
\resizebox{1.0\columnwidth}{!}{%
  \includegraphics{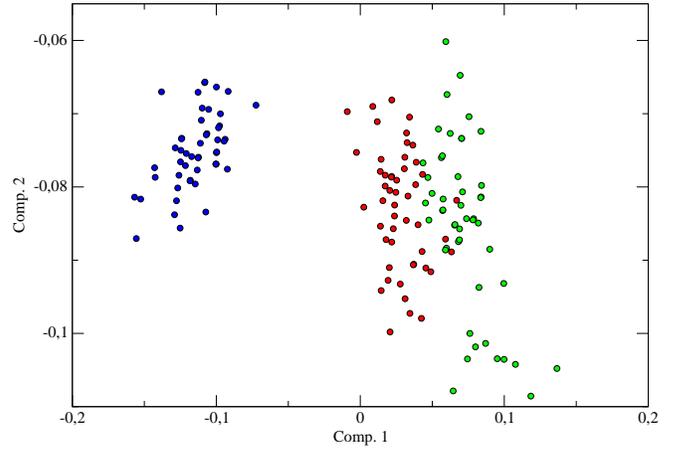}
}
\caption{Two principal components of the PCA analysis on the Iris dataset. Colors correspondence are: setosa-blue, versicolor-red, and virginica-green. The separation of pattern classes seems more clear in this projection.}
\label{pca}
\end{figure}

Following the steps of data clustering explained above, \CANVIS{we first performed a feature extraction$/$selection process. The idea here is simply to follow the workflow in any clustering problem, where the high dimensionality of the data and its redundancy is a main concern. In the particular case we analyze, we can use all the original data with no computational stress, however we propose to address the feature extraction using PCA which will be the most common approach in many scenarios. We performed the}
principal component analysis of the four features that form each pattern, and choose to work with the two principal components corresponding to the largest part of the data variance. In Fig.~\ref{pca} a representation of these two components is shown.
Based on these two variables, we propose to build up a similarity matrix as the euclidean distances between patterns components with respect to the center of mass of the data set in this space. For any pair of flowers $i$ and $j$, we define the similarity $s_{ij}={\bar d} - \| x^{i}-x^{j} \|)$, where ${\bar d}$ stands for the average distance of the set, and $\| \cdot \|$ is the euclidean distance between the feature vectors of each flower. The resulting similarity matrix is interpreted as a weighted network whose communities will, in principle, reproduce the right clustering of the data.

\begin{figure}[t]
\centering
\resizebox{1.0\columnwidth}{!}{%
  \includegraphics*{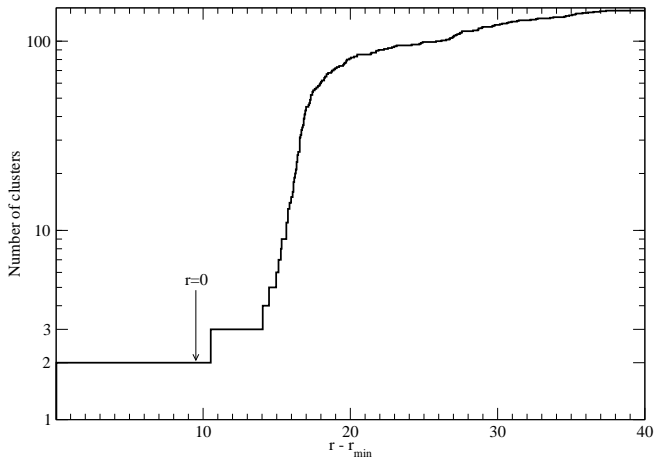}
}
\caption{Number of clusters as a function of the resolution parameter of the classification method (see text for details).}
\label{fig:plateaus}
\end{figure}

The results of the multiple resolution algorithm on the two main components of the Iris dataset is shown in the Fig.~\ref{fig:plateaus}. It can be observed that the longest plateau in terms of the resistance interval values is that formed by those partitions that divide the dataset into two communities. This is not a surprising fact, as we know beforehand that one of the three classes of flowers is linearly separable, and then this partition makes totally sense, since there is one for the Setosa class and the other one containing the Versicolor and Virginica. However, the second longest plateau is the one formed by the three community partitions, and if we analyze the most resistant of them, we realize that it largely corresponds to the biological taxonomy of the flowers. To be specific, if we calculate the success as the number of correctly classified nodes divided by the total number of nodes, we achieve for the most resistant partition of three communities a 94,6\% of success compared to the correct biological taxonomy.

Summarizing, we have presented a possible application of the multiple resolution method to the problem of data clustering. Our proposal has been proved competitive in success with other techniques used in the literature on the same benchmark \cite{pal}, but as an essential difference we also provide information of grouping at different scales of resolution that are invisible to other algorithms. The methodology presented so far is plausible to be extensive to any data clustering problem expressed in terms of similarity matrices.

\section{Conclusions}

Scientists working on the field of complex networks have developed tools for the analysis of structural information embedded in the topological connectivity matrix. Specially interesting are the heuristic algorithms intended to find the community structure of networks, which remind the kind of problems of data clustering found in many disciplines.Here we have presented a possible application of community detection algorithms to help exploratory analysis and data clustering. In particular, we have used a previous methodology proposed by the authors that allows for a multiple resolution of topological scales in the substructure of networks.

The exploratory analysis of the neural connectivity of the nematode {\em C.~elegans} has been presented. We found a tentative classification of groups of neurons presumably involved in specific tasks, according to the persistence of these groups in the topological analysis.
We have also exposed the applicability of the method to the unsupervised classification of data, using the famous Iris dataset as a benchmark. The results are encouraging, we observe the full spectrum of clusters according to the organization of data, and the most persistent scales are those corresponding to well-known facts about its structure, a partition in two linearly separable groups, and a partition in three groups corresponding to the biological taxonomy. These results open the field of applicability of the theory of complex networks to other problems where the representation of data as a network allows the use of the technology developed so far.

\appendix
\section{Functional groups of {\em C.~elegans}}

Classification of functional groups of neurons resulting from the multiple resolution method.
Using the database WormAtlas\cite{wa} and the results depicted in Fig.~\ref{freqtr} we have
identified nine groups of neurons of size lower than ten, whose functionality can be tentatively
related to a specific action. The process to assign a tentative function to the groups of neurons
has been done manually, reading the associated literature and using the worm-atlas database.
We expose the list in Table~\ref{celegansgroups}.

\begin{table}[th]
	\caption{\CANVIS{Temptative functionality of several significant groups of neurons found in the mesoscale.}}
	\begin{ruledtabular}
	\begin{tabular}{p{2.8cm}p{5cm}}
	Cluster~of~neurons & Tentative function \\ \hline\hline
RIAL, RIAR, RMDR, RMDVR, SMDVR, RMDDL, SMDDR & Nose/Head orientation movement.\\  \hline
IL1DR, IL1VR, IL2DR, IL2VR, RIPR & Head-withdrawal reflex, more related to dorsal relaxation. When worms are touched on either the dorsal or ventral sides of their nose with an eyelash, they interrupt the normal pattern of foraging and undergo an aversive head-withdrawal reflex.\\  \hline
IL2, IL2R, OLQVL, OLQVR, RIH & Head-withdrawal reflex, more related to ventral relaxation.\\ \hline
ADLR, AIBR, ASEL, ASHR, AWCL, AWCR, AIAR, AIYL & Olfactory and thermosensation reflex. \\ \hline
ASGL, ASJL, ASKL, AIAL, PVQL & Chemotaxis to lysine reflex. \\ \hline
DB1, DB2, DD1, VB2, VD2, AS3, DA2, DA3, DA4, DA5 & Backward sinusoidal  movement of the worm, more related to touch stimulus. \\ \hline
AVAL, AVAR, AVBL, AVBR, AVDL, AVDR, AVEL, AVER, DA1, FLPL &  Forward and Backward sinusoidal movement of the worm, more related to search for food in starving case, involve social feeding effect. \\ \hline
AVHL, AVHR, AVJL, AVFL, AVFR & Impossible to determine from the experimental data available. There is not any specific function known for any of these neurons. \\ \hline
AVKL, ACKR, PDEL, PDER, PVM, DVA, WN &  The functionality of this group could be related to a relaxation state similar to a sleep state, with reduced motor activity, decreased sensory threshold, characteristic posture and easy reversibility, basically mediated by  PDs neurons.
	\end{tabular}
	\end{ruledtabular}
	\label{celegansgroups}
\end{table}

\begin{acknowledgments}
We acknowledge support from the Spanish Ministry of Science and Technology FIS2009-13730-C02-02 and the Generalitat de Catalunya SGR-00838-2009.
\end{acknowledgments}


\end{document}